\documentclass[aps,prl,twocolumn,showpacs,groupedaddress]{revtex4}
\usepackage{graphicx}  
\usepackage{epsfig}
\usepackage{dcolumn}   
\usepackage{bm}        
\usepackage{amssymb}   
\begin{document}

\def\ra{{\rightarrow}}
\def\a{{\alpha}}
\def\b{{\beta}}
\def\l{{\lambda}}
\def\eps{{\epsilon}}
\def\T{{\Theta}}
\def\t{{\theta}}
\def\co{{\cal O}}
\def\car{{\cal R}}
\def\caf{{\cal F}}
\def\cs{{\Theta_S}}
\def\pr{{\partial}}
\def\tri{{\triangle}}
\def\na{{\nabla }}
\def\S{{\Sigma}}
\def\s{{\sigma}}
\def\sp{\vspace{.05in}}
\def\hs{\hspace{.25in}}
\def\rs{\vspace{-.08in}}

\newcommand{\be}{\begin{equation}} \newcommand{\ee}{\end{equation}}
\newcommand{\bea}{\begin{eqnarray}}\newcommand{\eea}
{\end{eqnarray}}


\begin{titlepage}

\title{\Large\bf Mass generation from a non-perturbative correction:\\
Massive NS-field and graviton in $(3+1)$-dimensions}

\author{Supriya Kar and R. Nitish}
\affiliation{Department of Physics \& Astrophysics, University of Delhi, New Delhi 110 007, India}

\begin{abstract}
We show that the massless form fields, in $(4+1)$-dimensional non-perturbation theory of emergent gravity, become massive in a perturbative phase without Higgs mechanism. In particular an axionic scalar sourced by a non-perturbative dynamical correction is absorbed by the form fields to describe a massive NS field theory on an emergent gravitational pair of $(3{\bar 3})$-brane. Arguably the novel idea of Higgs mechanism is naturally invoked in an emergent gravity underlying a ${\rm CFT}_6$. Analysis reveals ``gravito-weak'' and ``electro-weak'' phases respectively on a vacuum pair in $(4+1)$ and $(3+1)$-dimensions. It is argued that the massive NS field quanta may govern an emergent graviton on a gravitational $3$-brane.

\end{abstract}

\maketitle
\end{titlepage}

\noindent
{\it\bf Introduction:} The general theory of relativity (GTR) is governed by a metric tensor dynamics in 4D underlying a Pseudo-Riemannian manifold. The GTR is a second order formulation and is geometric. It describes an interacting classical theory and hence it rules out the possibility for a perturbation theory. Furthermore the coupled nature of differential field equations in GTR ensures non-linear solutions which are believed to be sourced by a non-linear energy-momentum tensor. Thus the quantum field dynamical correction to GTR urges for a  non-perturbation (NP) formulation in second order. 

\sp
\noindent
Interestingly the theoretical requirement has been attempted with a dynamical geometric torsion ${\cal H}_3$ in a second order while keeping the Neveu-Schwarz (NS) form onshell in an emergent first order \cite{abhishek-JHEP,abhishek-PRD,abhishek-NPB-P}. The non-perturbative formulation in a gauge choice has led to an emergent metric which turns out to be dynamical. Generically a geometric torsion in an emergent gravity is a dynamical formulation in $1.5$ order, where the metric dynamics can gain its significance at the expense of the non-perturbative dynamical correction \cite{kar-PRD-R}. The idea has led to a non-supersymmetric formulation for a NP-theory of quantum gravity in $(4+1)$-dimensions which may be identified with a stabilized string vacuum on a gravitational pair of $(3{\bar 3})$-brane. In addition need for an extra dimension to the GTR in a NP-theory of gravity is consistent with a fact that a ten dimensional type IIA superstring provides a hint towards a supersymmetric non-perturbation $M$-theory in eleven dimensions \cite{witten-NPB,schwarz-PLB}.

\sp
\noindent
In the article we present an elegant tool to generate mass for a gauge field by a geometric torsion in a $(4+1)$-dimensional NP-theory. Generically the NP-tool has been shown to generate a mass for the Neveu-Schwaz (NS) two form on a gravitational pair of $(3{\bar 3})$-brane. In particular a $(3+1)$ dimensional massive NS field quantum dynamics is argued to describe an emergent graviton in the same space-time dimension. It is shown that the local degree of the NP-correction is absorbed by the NS field and hence the axionic scalar in the NP-sector may formally been identified with a goldstone boson established in Higgs mechanism \cite{griffths}. Furthermore the emergent NP-theory, underlying a ${\rm CFT}_6$, is revisited with a renewed interest to reveal the Higgs mechanism naturally on a gravitational pair of $(4{\bar 4})$-brane. The emergent $(4+1)$ dimensional curvatures are argued to describe the ``gravito-weak'' phase of the NP-theory underlying the gravitational and weak interactions respectively on a ${\bar 4}$-brane and 
on a $4$-brane. The NP-correction is exploited to realize a duality between a strongly coupled weak interactions and weakly coupled gravity with 
a cosmological constant.

\sp
\noindent
{\it\bf Glimpse at non-perturbative physics:} In the context Dirichlet $(D)$ brane in ten dimensional type IIA or IIB superstring theory is believed to be a potential candidate to describe a non-perturbative world due to their Ramond-Ramond (RR) charges \cite{polchinski}. The $D$-brane dynamics is precisely governed by an open string boundary fluctuations and the Einstein gravity underlying a closed string is known to decouple from a $D$-brane. Interestingly for a constant  Neveu-Schwarz (NS) background in an open string theory, the $U(1)$ gauge field turns out to be non-linear on a $D$-brane and has been shown to describe an open string metric \cite{seiberg-witten}. The non-linear gauge dynamics on a $D$-brane is approximated by the Dirac-Born-Infeld (DBI) action. Various near horizon black holes have been explored using the open string metric on a $D$-brane in the recent past  \cite{gibbons,mars,ishibashi,kar-majumdar,kar-PRD,kar-JHEP,liu,zhang2}. 

\sp
\noindent
However the mathematical difficulties donot allow an arbitrary NS field to couple to an open string boundary though it is known to describe a torsion in ten dimensions. A torsion is shown to modify the covariant derivative and hence the effective curvatures in a superstring theory \cite{candelasHS,freed}. In the recent past a constant NS field on a $D_4$-brane has been exploited for its gauge dynamics in an emergent theory \cite{abhishek-JHEP,abhishek-PRD,abhishek-NPB-P}. In particular the Kalb-Ramond (KR) field dynamics are used to define a modified derivative ${\cal D}_{\mu}$ uniquely. It has been shown to govern an emergent curvatures on a gravitational pair of $(3{\bar 3})$-brane.

\sp
\noindent
The stringy pair production by the KR form primarily generalizes the established Schwinger pair production mechanism \cite{schwinger}. The non-perturbation tool was vital to explain the Hawking radiation phenomenon \cite{hawking} at the event horizon of a black hole. The novel idea was applied to the open strings pair production \cite{bachas-porrati} by an electromagnetic field. Furthermore the mechanism was explored to argue for the $M$-theory underlying a vacuum creation of $(D{\bar D})_9$ pair at the cosmological horizon \cite{majumdar-davis}. 

\sp
\noindent
In particular the stringy pair production by the KR quanta has been explored in diversified contexts to obtain:
(i) a degenerate Kerr \cite{sunita-NPB,sunita-IJMPA}, (ii) a natural explanation to quintessential cosmology \cite{priyabrat-EPJC,priyabrat-IJMPA,priyabrat-IJIRSET,deobrat-IJIRSET}, (iii) a emergent Schwarzschild/topological de Sitter, $i.e.$ a mass pair on $(4{\bar 4})$-brane \cite{richa-IJMPD,deobrat-Springer} and (iv) a fundamental theory in twelve dimensions and an emergent $M$-theory in eleven dimensions \cite{kar-PRD-R}. Generically the {\it stringy} nature and the {\it pair production tool} respectively ensure a quantum gravity phase and a non-perturbative phenomenon. Thus an emergent stringy pair is believed to describe a NP-theory of emergent gravity in $1.5$ order formulation. Preliminary investigation has revealed that the NP-theory sourced by a ${\rm CFT}_6$ may lead to an unified description of all four fundamental forces 
in nature. Analysis is in progress and is beyond the scope of this article.

\sp
\noindent
{\it\bf Two form (KR${\mathbf{\leftrightarrow}}$NS) dynamics:}
We begin with the KR form $U(1)$ dynamics on a $D_4$-brane in presence of a background  (open string) metric $G^{(NS)}_{\mu\nu}$ which is known to be sourced by a constant NS form \cite{seiberg-witten}. The gauge theoretic action is given by
\rs
\be
S={{-1}\over{(8\pi^3g_s){\alpha'}^{3/2}}}\int d^5x {\sqrt{-G^{({\rm NS})}}}\ H_{\mu\nu\lambda}H^{\mu\nu\lambda}\ ,\label{gauge-2}
\ee

\rs
\noindent
where $G^{(NS)}$$=$$\det\ G^{(NS)}_{\mu\nu}$. The KR field dynamics $H_3$ is absorbed, as a torsion connection, and modifies $\nabla_{\mu}\rightarrow {\cal D}_{\mu}$. The modified derivative leads to an emergent description where the NS field becomes dynamical \cite{abhishek-JHEP}. It defines a geometric torsion: 
\rs
\bea
&&{\cal H}_{\mu\nu\lambda}=\ {\cal D}_{\mu}B^{(NS)}_{\nu\lambda}+\ {\rm cyclic\ in}\ (\mu,\nu,\lambda)\ ,\nonumber\\
&&\quad =H_{\mu\nu\rho}B_{\lambda}^{(NS)\rho}+ H_{\mu\nu\alpha}B_{\rho}^{(NS)\alpha} B_{\lambda}^{(NS)\rho}+\dots\label{gtorsion-1}
\eea

\rs
\noindent
The $U(1)$ gauge invariance of ${\cal H}_3^2$ under NS field transformation incorporates a symmetric $f_{\mu\nu}={\bar{\cal H}}_{\mu\alpha\beta}{{\cal H}^{\alpha\beta}{}}_{\nu}$ correction which in turn defines an emergent metric: $G^{\rm EG}_{\mu\nu}= G^{(NS)} \pm f_{\mu\nu}$. The generic curvature tensors are worked out using the commutator of the modified derivative operator:
\rs
\bea
&&\left [ {\cal D}_{\mu},{\cal D}_{\nu}\right ]A_{\lambda}= \left ({{\cal R}_{\mu\nu\lambda}{}}^{\rho}+
{{\cal K}_{\mu\nu\lambda}{}}^{\rho}\right )A_{\rho}- 2{{\cal H}_{\mu\nu}{}}^{\rho}\ {\cal D}_{\rho}A_{\lambda}\ ,\nonumber\\
&&\left [ {\cal D}_{\mu},{\cal D}_{\nu}\right ]\psi= -2\ {{\cal H}_{\mu\nu}{}}^{\rho}\ {\cal D}_{\rho}\psi\ ,\label{new-curvature}
\eea

\rs
\noindent
where ${{\cal R}_{\mu\nu\lambda}{}}^{\rho}$ denotes the Riemann tensor. For a constant metric the Riemann tensor becomes trivial.
${\cal H}_3$ ensures a NS field dynamics in an emergent metric scenario. The fourth order curvature tensor ${\cal K}_{\mu\nu\lambda\rho}$ 
can be splitted into a pair symmetric and a pair non-symmetric under an interchange of first and second pair of indices. The irreducible curvatures have been worked out \cite{kar-PRD-R} to obtain emergent NP-theory of gravity for onshell NS field.

\sp
\noindent
{\bf Mass generation as a non-perturbation effect:}
We begin with a NP-theory of emergent gravity in $(4+1)$-dimensions underlying a geometric torsion ${\cal H}_3$ in $1.5$ order formulation \cite{kar-PRD-R}. The effective action has been shown to govern a NS field dynamics in an emergent first order (perturbation) gauge theory and a local geometric torison ${\cal H}_3$ in a second order NP-theory. It is given by
\rs
\bea
S_{\rm NP}&=&{1\over{\kappa'}^3}\int d^5x\ {\sqrt{-g}}\ \Big ({\cal K}\ -\ {1\over{48}}\ {\cal F}_4^2\Big )\ ,\nonumber\\
{\rm where}\quad {{\cal F}}_4&=&{\sqrt{2\pi\alpha'}}\big ( 
d{\cal H}_3-{\cal H}_3\wedge {\cal F}_1\big )\ ,\label{NG5-main}
\eea

\rs
\noindent
Equivalently the emergent theory may be described by the geometric form(s). We set ${\kappa'}^2=(2\pi\alpha')=1$ in the article. Then the effective actions are:
\rs
\bea 
S_{\rm NP}&=&-\ {1\over{12}}\int\ {\sqrt{-g}}\ \Big ({\cal H}_{\mu\nu\lambda}{\cal H}^{\mu\nu\lambda}+6\big ({\cal D}_{\mu}\psi\big )\big ({\cal D}^{\mu}\psi\big )\Big )\ , \nonumber\\
&=&-\ {1\over{4}}\int {\sqrt{-g}}\Big ({\cal F}_{\mu\nu}{\cal F}^{\mu\nu}+ 2\big ({\cal D}_{\mu}\psi\big )\big ({\cal D}^{\mu}\psi\big )\Big ) \ . \label{NG5-1}
\eea

\rs
\noindent
The first term, in all three actions (\ref{NG5-main})-(\ref{NG5-1}), sources an emergent metric and hence a torsion free geometry in absence of the second term there. A propagating geometric torsion is described by the second term which is indeed a dynamical NP-correction. The emergent curvature scalar ${\cal K}$ and its equivalent Lorentz scalars constructed from the geometric forms ${\cal H}_3$ and ${\cal F}_2$ can govern an emergent metric. Each of them possess three local degrees in an emergent first order formulation. The ${\cal F}_4$ is Poincare dual to a dynamical axionic scalar field $\psi$ and possesses one local degree in an emergent second order formulation. Together they describe four local degrees in a NP-theory of emergent theory of gravity in $5D$. In addition the NP-formulation is described by an appropriate toplogical coupling from: 

\vspace{-.06in}
\noindent
$$\Big ( B_2^{(KR)}\wedge{\cal H}_3\ ,\;\; B^{(NS)}_2\wedge H_3\ ,\;\; B_2^{(NS)}\wedge{\cal F}_2\wedge d\psi\Big )\ .$$

\vspace{-.06in}
\noindent
A geometric ${\cal F}_2$ in an emergent theory underlies the $U(1)$ gauge symmetry and is given by
\rs
\be
{\cal F}_{\mu\nu}=\Big ({\cal D}_{\mu}A_{\nu} -{\cal D}_{\nu}A_{\mu}\Big )
=\Big (F_{\mu\nu}+ {{\cal H}_{\mu\nu}{}}^{\lambda}A_{\lambda}\Big )\ ,\label{NG5-2}
\ee

\rs
\noindent
where $F_{\mu\nu}=\big (\nabla_{\mu}A_{\nu}-\nabla_{\nu}A_{\mu}\big )$. The geometric forms ${\cal F}_2$ and ${\cal H}_3$ are are worked out for their gauge theoretic counter-parts. The Lorentz scalar for a geometric two form may be re-expressed with a mass (squared) matrix represented by a symmetric (emergent) curvature tensor of order two. It is given by 
\rs
\be
{\cal F}_{\mu\nu}^2=\Big (F_{\mu\nu}^2 -{\cal K}^{\mu\nu}A_{\mu}A_{\nu}  + {{\epsilon^{\mu\nu\lambda\alpha\beta}}\over{\sqrt{-g}}}A_{\mu}F_{\nu\lambda}{\cal F}_{\alpha\beta}\Big )\ ,\label{NG5-3}
\ee

\rs
\noindent
where the symmetric curvature tensor of order two may be expressed in terms of a geometric 3-form and its Poincare dual. They are:
\be
{\cal K}^{\mu\nu}=-{1\over4} {\cal H}^{\mu\alpha\beta}{{\cal H}_{\alpha\beta}{}}^{\nu}=\Big (g^{\mu\nu}{\cal F}^2_2 + 2{\cal F}^{\mu\lambda}{{\cal F}_{\lambda}{}}^{\nu}\Big )\ .\label{NG5-4}
\ee
The geometric two form in an emergent nonperturbation theory (\ref{NG5-1}) is replaced by the gauge theoretic forms (\ref{NG5-3}). A priori the effective non-perturbative dynamics is re-expressed as:
\rs
\bea
S_{\rm NG}&=&-\ {1\over{4}}\int {\sqrt{-g}}\ \Big [ F_{\mu\nu}^2 -{\cal K}^{\mu\nu}A_{\mu}A_{\nu} + 2\big (\nabla_{\mu}\psi\big )^2\Big ]\nonumber\\
&&+\int \;\; \Big (A_1\wedge F_2\wedge F_2 -\ B_2^{(NS)}\wedge H_3\Big )\ . \label{NG5-5}
\eea

\rs
\noindent
At a first sight the emergent curvature tensor ${\cal K}^{\mu\nu}$ appears to a mass (squared) matrix.
A count for the local local degrees enforces ${\cal F}_4=0$ in the effective gauge theory (\ref{NG5-5}). Thus a geometric torsion turns out to be a constant which in turn defines a perturbative vacuum. However ${\cal F}_4\neq 0$ in an emergent gravity (\ref{NG5-4}) turns out to be non-trivial. Alternately the perturbative gauge vacuum may be realized in a gauge choice for ${\cal F}_4=0$. A constant ${\cal H}_3$ leads to a constant ${\cal K}^{\mu\nu}$ which is diagonalized. Thus ${\cal K}^{\mu\nu}$ is proportional to $g^{\mu\nu}$ in a perturbation theory:
\rs
\be
{\cal K}^{\mu\nu}=\ m_1^2\ g^{\mu\nu}=\ {1\over5}g^{\mu\nu}{\cal K}\ ,\label{NG5-6}
\ee

\rs

\rs
\noindent
where $m_1^2$ is a proportionality constant. It assigns a mass to $A_{\mu}$ at the expense of a dynamical non-perturbative correction. Interestingly the non-perturbative tool to generate a mass for a gauge field is remarkable. In fact it helps to generate mass $m_p={\sqrt{{\cal K}/d}}$ for a generic higher $p$-form field in a gauge theory in $d$-dimensions. Furthermore a mass $m_1$ can also be derived from a geometric two form in an appropriate combination (\ref{NG5-4}). With a proportionality constant ${\tilde m}_1^2$: the symmetric tensor ${\cal K}^{\mu\nu}={\tilde m}_1^2g^{\mu\nu}$ and the curvature scalar ${\cal K}=3{\cal F}_{\mu\nu}^2$. Then the mass ${\tilde m}_1$ for $A_{\mu}$ field is re-expressed generically in $d$-dimensions. It is given by
\rs
\be
{\tilde m_1}^2=\ {{d-2}\over{d}} \ {\cal F}_{\mu\nu}^2\ .\label{NG5-7}
\ee

\rs
\noindent
It can be checked that ${\tilde m}_1=m_1$ and hence the mass of an one form is uniquely defined in a perturbative gauge theory using a NP-technique. The Poincare dual of four form ensures that the local degree of an axionic scalar signifying a NP-dynamics is absorbed to generate a massive gauge field in a perturbation gauge theory which is equivalently described by a massless gauge field in a NP-theory of emergent gravity.

\sp
\noindent
The correspondence signifies a strong-weak coupling duality symmetry \cite{ashoke-IJMPA} in an emergent gravity underlying a strongly coupled 
NP-theory and a weakly coupled perturbation theory. Interestingly the axion in a NP-theory may be identified with a goldstone boson in a spontaneous local $U(1)$ symmetry breaking phase of a peturbative vacuum. Then the effective action (\ref{NG5-1}) in a weakly coupled gauge theory may formally be re-expressed as:
\rs
\bea
S_{\rm PG}&=&-\ {1\over4}\int d^5x {\sqrt{-G}}\ \Big (F^2_2 - m_1^2 A^2\Big )\nonumber\\
&-&\int \Big (A_1\wedge F_2\wedge F_2 + B_2^{(KR)}\wedge {\cal H}_3\Big )\ ,\label{NG5-51}
\eea

\rs
\noindent
where ${\cal F}_2\rightarrow F_2$ in a perturbation gauge theory. Importantly a massless gauge field $A_{\mu}$ in an emergent non-perturbation theory of gravity in $5D$ becomes massive at the expense of a non-perturbative dynamics. The non-perturbative tool for mass generation of a gauge field in a perturbative vacuum is remarkable and appears to be a generic feature for higher forms. It is believed to be a viable NP-tool to explore new physics underlying a strong-weak coupling duality. A massive gauge field dynamics for its Poincare dual is worked out to assign a mass $m_2$ to the KR field in a perturbative gauge theory. Computation of mass (squared) matrix for a NS field may directly be worked out from the curvature scalar:
\rs
\be
{\cal K}\approx-{1\over4}\Big ({H^{\lambda}{}}_{\alpha\beta}{H^{\alpha\beta}{}}_{\rho}\Big ) B^{(NS)}_{\delta\lambda}B_{(NS)}^{\delta\rho}\ .\label{NG5-52}
\ee

\rs
\noindent
In a gauge choice for a nonpropagating geometric torsion, the gauge theoretic $H_3$ turns out to be a constant for a perturbative vacuum within a non-perturbative formulation. This is due to a fact that the NS field is covariantly constant on a $D_4$-brane where $\nabla_{\mu}$ is an appropriate covariant derivative. Thus the mass (squared) matrix for the NS field in eq(\ref{NG5-52}) can be diagonal and hence is proportional to $g^{\lambda}_{\rho}$. It implies
\rs
\be
\Big ({H^{\lambda}{}}_{\alpha\beta}{H^{\alpha\beta}{}}_{\rho}\Big )\ = m_2^2\ g^{\lambda}_{\rho}\ .\label{NG5-53}
\ee

\rs
\noindent
At this juncture we recall a transition from the KR gauge theory on a $D_4$-brane 
defined with a constant NS background with that of a NP-formulation of an emergent gravity on gravitational pair of $(3{\bar 3})$-brane \cite{abhishek-JHEP,abhishek-PRD}. Generically it underlies a correspondence between a non-perturbation emergent gravity on a gravitational pair of $(4{\bar 4})$-brane and a perturbation CFT on a $D_5$-brane. The boundary/bulk correspondence ${\rm NP}_5/{\rm CFT}_6$ may be summarized with the relevant forms:
\be
\left [{\cal H}_3\ ,\ {\cal F}_4\ ,\ B_2^{(KR)}\right ]_{\rm NP}\nonumber\\
\longleftrightarrow\ \left [\ H_3\ ,\ B_2^{(NS)}\right ]_{\rm CFT}\ .\label{NG5-54}
\ee

\rs
\noindent
Primarily the dynamical correspondence is between a KR form in the world-volume gauge theory and a NS form in superstring theory. 
Eq(\ref{NG5-53}) further ensures that a mass for NS-field is sourced by the KR field dynamics. Similarly the analysis, followed from the derivation of an effective action (\ref{NG5-51}), confirms that a mass for KR field is indeed sourced by a NS field dynamics. 
Both of them are two forms and they are different due to their differences in backgrounds or connections. Intuitively the dynamical correspondence
(\ref{NG5-54}) leading to two different formulations may be viewed with a single two form with two different names for their masses in a perturbative vacuum.

\sp
\noindent
The dynamical correspondence between a perturbative gauge theory and a non-perturbative emergent gravity is remarkable. It signifies a strong/weak coupling duality \cite{ashoke-IJMPA} between the two different formulations underlying a two form gauge theory. The NP-theory of emergent gravity is purely governed by ${\cal H}_3$ and hence generically describes a torsion geometry. However in a gauge choice ${\cal F}_4=0$, the emergent gravity describes a torsion free geometry purely sourced by a dynamical NS field.

\sp
\noindent
A realization of perturbative vacuum (\ref{NG5-51}) within a NP-theory may be described for a KR field. It is given by 
\rs
\bea
S_{\rm PG}&=&-\ {1\over{12}}\int d^5x {\sqrt{-G}}\ \Big (H_{\mu\nu\lambda}H^{\mu\nu\lambda} - m_2^2 B^2_{(KR)}\Big )\nonumber\\
&&\quad\ -\ \int \Big ( \big ({\cal F}_2 - B_2^{(NS)}\big )\wedge H_3\Big )\ .\label{NG5-6}
\eea

\rs
\noindent
The first term in the bulk topological action is a total divergence. However it regains significance at the $4D$ boundary where the coupling $\big (B_2\wedge {\cal F}_2\big )$ may be identified with the $BF$ toplogical theory as discussed in refs\cite{BF-1,BF-2}. A massive KR form in a perturbation gauge theory is generated by a NP-correction sourced by a propagating geometric torsion which turns out to be an axion in $5D$. The 
NP-tool to generate mass for a form field underlying a geometric torsion in $1.5$ order formulation is thought provoking.
\begin{figure}
\centering
\mbox{\includegraphics[width=1.0\linewidth,height=0.28\textheight]{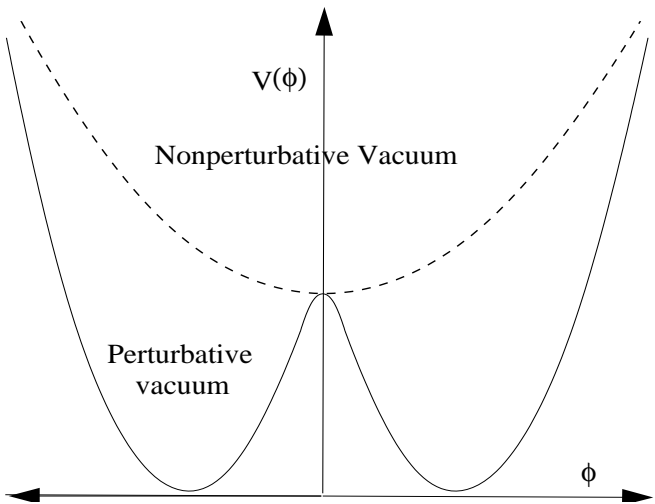}}
\caption{\it Potential variation shows that a non-perturbative stable vacuum may be viewed as a perturbative unstable vacuum}
\end{figure}

\noindent
The emergent gravity scenarios \cite{abhishek-JHEP,abhishek-PRD,abhishek-NPB-P,sunita-NPB,sunita-IJMPA,priyabrat-EPJC,priyabrat-IJMPA,priyabrat-IJIRSET,
deobrat-IJIRSET,richa-IJMPD,deobrat-Springer} ensure that all the NP-phenomena are sourced by a lower dimensional $D_p$-brane whose fundamental unit is a $D$-instanton. Interestingly, in a recent article \cite{kar-PRD-R}, the NP-phenomenon has been shown to be sourced by the dynamics of ${\cal H}_3$ potential in $1.5$ order formulation. Dynamical effect incorporates a quantum correction to the torsion free vacua underlying an emergent metric. The correction breaks the Riemannian geometry and hence is hidden to the GTR underlying an an emergent $3$-brane universe within a gravitational pair $(3{\bar 3})$-brane. 

\sp
\noindent
The NP-idea leading to mass generation suggests that a dynamical axion (quintessence) or generically a higher-essence is hidden to an emergent $3$-brane universe and hence its significance to the GTR can only be revealed with a topological coupling.

\sp
\noindent
{\bf Higgs mechanism in emergent gravity:} We begin by recalling the perspectives of a CFT underlying a KR gauge theory on a $D_5$-brane. The gauge theoretic vacuum may equivalently be described by an a prior massless NS form in an emergent $6D$ perturbation theory \cite{richa-IJMPD}. A pair-symmetric emergent curvature tensor of order four has been shown to be sourced by a NS field in an emergent (first order) perturbation theory and possesses six local degrees. It has been shown to describe a torsion free geometry and has been argued to describe a Riemann type curvature in $6D$. A dynamical correction by ${\tilde{\cal F}}_4^2$ in an emergent gravity theory incorporates four non-perturbative local degrees. The effective dynamics is described by ten local degrees and is given by
\rs
\be
S=\int d^6x\ {\sqrt{-{\tilde g}}}\ \Big ({\tilde{\cal K}} -{1\over{48}}\ {\tilde{\cal F}}_4^2\Big )\ .\label{NG6-1}
\ee

\rs
\noindent
Interestingly the local degrees of a NS field in $6D$ precisely match with the local degrees of a metric tensor in $5D$ and a scalar field presumably underlying a quintessence. Generically a two form ($NS$ field) theory in the bulk can completely be mapped to the boundary dynamics underlying a metric tensor and a scalar field $\phi$. The bulk/boundary correspondence in an emergent gravity formulation on a gravitational pair of $(4{\bar 4})$-brane is remarkable. It is believed to attribute Riemannian geometry possibly at the expense of a local $U(1)$ gauge symmetry. We digress to mention that
attempts have been made to use the gauge principle to realize Riemannian geometry in the recent past \cite{wilczek-PRL}. The effective action in the case is given by 
\rs
\bea
S&=&\int_{4{\bar 4}} d^5x\ {\sqrt{-G}}\ \Big ({\cal R}\ - {1\over4}{\cal F}_{\mu\nu}{\cal F}^{\mu\nu}\nonumber\\
&&\qquad\quad - \big ({\cal D}_{\mu}\Phi\big )^{\star}\big ({\cal D}^{\mu}\Phi\big )\ - V(\Phi,\Phi^{\star})\ \Big )\ .\label{NG6-2}
\eea

\rs
\noindent
The complex scalar field: $\Phi={1\over{\sqrt{2}}}\left ( \phi+i\psi \right )$ is defined with two real scalar fields where $\psi$ denotes an axionic scalar sourced by a NP-correction. A geometric two form in $5D$, though derived from a NP-dynamics in $6D$, describes a perturbative field strength. This is due to fact that the ${\cal H}_3$ dynamics can not be realized by ${\cal F}_2$ in $5D$. The canonical potential $V$ is sourced by the gravitational interaction in $6D$. An explicit form may be assigned to $V$ by taking an account for the self-interaction of the complex scalar field with a wrong sign for the mass term underlying an unstable perturbative vacuum. It may suggests that Higgs mechanism may find a natural place 
on a gravitational pair of $(4{\bar4})$-brane. The potential may explicitly be given by
\rs
\be
V(\Phi,\Phi^{\star})\ =\Big (m^2\big ({\Phi}^{\star}\Phi\big )-\lambda^2 \big (\Phi^{\star}\Phi\big )^2\Big )\ ,\label{NG6-3}
\ee

\rs
\noindent
where $m$ and $\lambda$ are real constants. The emergent theory (\ref{NG6-2}) with the potential (\ref{NG6-3}) remains invariant under a global $U(1)$ symmetry: $\Phi\rightarrow e^{i\theta}\Phi$. The global $U(1)$ is replaced with a local $U(1)$ symmetry with a minimal gauge coupling in the action: $D_{\mu}\equiv\big ({\cal D}_{\mu} +ieA_{\mu}\big )$.  Explicitly the 
dynamics leading to an unstable vacuum is given by
\rs
\bea
S&=&-\int_{4{\bar 4}}d^5x\ {\sqrt{-G}}\ \Big [\ {\cal R}\ - {1\over4}{\cal F}_{\mu\nu}^2 - \big (D_{\mu}\Phi\big )^{\star}\big (D^{\mu}\Phi\big )\nonumber\\
&&\qquad\qquad\qquad -\ m^2\big ({\Phi}^{\star}\Phi\big )+\lambda^2 \big (\Phi^{\star}\Phi\big )^2\ \Big ]\ .\label{NG6-31}
\eea

\rs
\noindent
The interaction energy function $V(\Phi^{\star},\Phi)$ at its minima satisfies an equation of a circle: 
\rs
$$\phi_{{\rm min}}^2+\psi_{{\rm min}}^2= \Big ( {m\over{\lambda}}\Big )^2\ .$$ 

\rs
\noindent
Thus a large number of stable ground/vacuum states, underlying the local $U(1)$ symmetry, are described by the circle equation. Any particular vacuum state, $i.e.\ \phi_{{\rm min}}=\big (m/\lambda\big )$ and $\psi_{{\rm min}}=0$, spontaneously breaks the local $U(1)$ symmetry in an emergent geometric theory. In fact the local symmetry breaking phenomenon, $i.e.$ Higgs mechanism, takes place at the event horizon of an emergent black hole which is identified as a stable vacuum.

\sp
\noindent
A shift from an unstable (a non-perturbation) vacuum (\ref{Higgs-1}) to a stable (perturbative) vacuum may be realized with redefined real scalar fields: $\eta=\phi-\big ( m/\lambda\big )$ and $\xi=0$. The action is re-expressed in terms of $\eta$ and $\xi$ fields for a stable vacuum and is known to generate mass term for the gauge field $A_{\mu}$ in addition to a few non-sensible interactions. For instance see a text book \cite{griffths} for the detailed nature of interactions in the symmetry breaking phase underlying the Higgs mechanism. The non-sensible interaction terms can be gauged away completely by the local $U(1)$ invariance under $\Phi\rightarrow\Phi'$ in the action (\ref{Higgs-1}). In a gauge choice: 
$\Phi'=\big (\phi\cos\theta -\psi\sin\theta\big )$, $i.e.$ restricting to the real parts, the complete perturbation theory on a gravitational pair is given by
\bea
&&S_{\rm PG}=\int_{4{\bar 4}}d^5x{\sqrt{-G}}\ \Big [ \Big ({\cal R} - {{m^4}\over{4\lambda^2}}\Big )+{{e^2}\over2} \Big (\eta+{{m}\over{\lambda}}\Big )^2A^2\nonumber\\
&&-{1\over4}{\cal F}_{\mu\nu}^2- {1\over2} \big ({\cal D}\eta\big )^2+ {1\over2}m^2\eta^2 +\big (m\lambda\big )\eta^3 + {1\over4}{\lambda^2\eta^4}\Big ]\ .\label{Higgs-1}
\eea
It implies that a cosmological constant appears to possess its origin in the symmetry breaking phase and is sourced by the Higgs mechanism.
Keeping a track for the four local degrees in $5D$ emergent gravity, the effective dynamics may explicitly be given on a $4$-brane within a vacuum pair of gravitational brane/anti-brane. Thus some of the undesirable emergent curvatures are assigned to an anti $4$-brane.  It is equivalent to a consistent truncation of the effective action defined with a massive gauge field. It may suggest that the analysis under a CFT leads to a study of Higgs mechanism naturally in an emergent gravity in $5D$. Then the effective dynamics on an emergent gravitational $4$-brane in presence of a background 
${\bar 4}$-brane is re-expressed as:
\bea
S_{\rm PG}&=&-{1\over4}\int_4 d^5x\ {\sqrt{-G}}\ \Big [{\cal F}_{\mu\nu}^2\ -\ {{e^2}\over2} \Big (\eta+{{m}\over{\lambda}}\Big )^2A^2\Big ]\nonumber\\
&&+\int_{\bar 4}d^5x\ {\sqrt{-G}}\ \Big [\ \Big ({\cal R}\ - {{m^4}\over{4\lambda^2}}\Big )\ - {1\over2} \big ({\cal D}\eta\big )^2
\nonumber\\
&&\qquad\ + \ {1\over2}m^2\eta^2 + \big (m\lambda\big )\eta^3 + {1\over4}{\lambda^2\eta^4}\ \Big ]
\ .\label{Higgs-1}
\eea 
The gauge field on an emergent $4$-brane universe acquires a mass $M={e\over{\sqrt{2}}}\Big (\eta_0+{{m}\over{\lambda}}\Big )$ via Higgs mechanism where the Higgs field takes a constant $\eta_0$ there. Apparently the local degree of the self interacting Higgs scalar is described on an anti $4$-brane and is hidden to the $4$-brane-universe. Thus the Higgs field, underlying a NP-formulation of emergent gravity, may be identified with a missing scalar in a $5D$ metric theory. It is inspiring to interpret the Higgs scalar as a hidden-essence to the gravitation theory in $5D$. The scalar field, being a generalized coordinate, determines the thickness of the brane/anti-brane configuration. 

\sp
\noindent
In a NP-decoupling limit a gravitational $4$-brane becomes independent from the anti $4$-brane and hence $\eta\rightarrow \eta_0$. The effective dynamics of a $4$-brane and anti $4$-brane are approximated in the limit to yield:
\bea
&&S_{4}\rightarrow -{1\over{12}}\int\ d^5x\ {\sqrt{-G}}\ \Big ({\cal H}_{\mu\nu\lambda}^2\ -{\tilde M}^2 B_{(NS)}^2\Big )\nonumber\\
{\rm and}&&S_{\bar 4}\rightarrow \int\ d^5x\ {\sqrt{-G}}\ \Big ({\cal R}-\Lambda  \Big )\ ,\label{Higgs-5D-6}
\eea
where $\Lambda= (m^4/4\lambda^2)$ is a constant. A mass for a NS field ensures a short range interactions. Thus the effective dynamics on a $4$-brane may be identified with a weak interacting phase for the NS boson in a decoupling limit. Remarkably an anti $4$-brane effective dynamics is purely governed by the Riemannian geometry in the limit. Generically the action (\ref{Higgs-1}) signals a ``gravito-weak'' phase within a NP-theory.

\sp
\noindent
Furthermore the emergent gravity on a $4$-brane may further be viewed on a gravitational pair of $(3{\bar 3})$-brane. The effective action is given by
\rs
\bea
S_{\rm PG}&=&-{1\over{12}}\int_{3} d^4x {\sqrt{-G}}\ \Big ({\cal H}_3^2\ -{\tilde M}^2 B_{(NS)}^2\Big )\nonumber\\
&&\; -\ {1\over{4}}\int_{{\bar 3}} {\sqrt{-G}}\ {\cal F}_2^2-\int_{3{\bar 3}}\ B_2^{(NS)}\wedge {\cal F}_2\
\ .\label{Higgs-5D-7}
\eea
Four local degrees in $5D$ may rightfully be governed by two local degrees of a massive NS field on an emergent  $3$-brane and two for a 
massless gauge field $A_{\mu}$ on an anti $3$-brane. This is due to a fact that GTR and their parallel are described by two local degrees each 
in an emergent scenario. Two local degrees of a massive NS field in $(3+1)$-dimensions is a NP-phenomenon as the mass is generated by a NP-local 
degree in $5D$. The correspondence between the NS-field in $5D$ and a metric field in $4D$ with a quintessence scalar further re-confirms 
two local degrees of a massive NS field in $(3+1)$-dimensions. It suggests that the massive NS field quanta in $(3+1)$-dimensions with a hidden 
NP-axion may be a potential candidate to describe a graviton in $4D$.

\sp
\noindent
A mass for a NS field in the action (\ref{Higgs-5D-7}) ensures that an emergent $3$-brane may formally be identified with the weak interacting NS boson, whose role is analogous to the gauge bosons $(W^{\pm}, Z^0)$ in standard model for particle physics. The dynamics on an anti $3$-brane governs an $U(1)$ gauge theory and may be identified with an EM-vacuum. Remarkably the complete dynamics (\ref{Higgs-5D-7}) may a priori be viewed via  ``electro-weak'' interactions. 

\sp
\noindent
On the other hand the topological term in eq(\ref{Higgs-5D-7}) precisely describes a coupling between an emergent gravitational $3$-brane and an anti $3$-brane within a vacuum pair. Generically an emergent gravity on a $3$-brane underlying a Riemann curvature may be derived from the anti $4$-brane (\ref{Higgs-5D-6}). In a decoupling limit the quintessence freezes to describe the GTR. For constant values: $\eta_1>\eta_0>\eta_{-1}$,  $i.e.$ for $\eta_1= (1.707)\eta_0$ and $\eta_{-1}= (0.293) \eta_0)$, 
the non-perturbation correction decouples to yield: 
\rs
\be
S_{3}=\int\ d^4x\ {\sqrt{-G}}\ \Big ({\cal R}-\Lambda_{\rm eff}  \Big )\ ,\label{Higgs-5D-8}
\ee

\vspace{-.35in}
$${\rm where}\;\ \Lambda_{\rm eff}=\Lambda\ - {{\eta_o^2}\over{2}}\Big [\big (m+\ \lambda\eta_1\big )\big (m+\ \lambda\eta_{-1}\big )\Big ]\ .$$

\vspace{-.1in}
\noindent
Thus the Higgs scalar in a NP-decoupling limit ensures a small cosmological constant. 
Analysis suggests that the Einstein-Hilbert action in presence of a small non-zero value for $\Lambda_{\rm eff}$ may alternately be realized by the Higgs phase of NS field presumably underlying a ``gravito-weak'' interaction on a gravitational pair of $(3{\bar 3})$-brane. The Higgs scalar $\eta$ and the scalar derived from ${\cal R}$ in eq(\ref{Higgs-5D-6}) are identified as the quintessence(s) for two emergent pairs of $4D$ brane dynamics. Presumably it provides a hint towards four parallel brane-universes in $4D$ underlying a NP-theory in $6D$. The unification idea \cite{wilczek-Unification} underlying a two form CFT is thought provoking and is believed to reveal new physics.

\vspace{.1in}

\sp
\noindent
{\it\bf Acknowledgements:}
Author (SK) acknowledges the research and development grant by the University of Delhi, India. A preliminary version of the research was presented by the author (RN) in an International Conference on ``New Trends in Field Theories 2016 November 06-10'' at the 
Banaras Hindu University, Varanasi, India. Authors gratefully acknowledge various discussions in general during the conference.

\def\anp{Ann. of Phys.}
\def\cmp{Comm.Math.Phys.\ {}} {}
\def\springer{Springer.Proc.Phys.}
\def\prl{Phys. Rev. Lett.}
\def\prd#1{{Phys.Rev.} {\bf D#1}}
\def\jhep{JHEP\ {}}{}
\def\cqg{Class.\& Quant. Grav.}
\def\plb#1{{Phys. Lett.} {\bf B#1}}
\def\npb#1{{Nucl. Phys.} {\bf B#1}}
\def\mpl#1{{Mod. Phys. Lett} {\bf A#1}}
\def\ijmpa#1{{Int.J.Mod.Phys.} {\bf A#1}}
\def\ijmpd#1{{Int.J.Mod.Phys.} {\bf D#1}}
\def\mpla#1{{Mod.Phys.Lett.} {\bf A#1}}
\def\rmp#1{{Rev.Mod.Phys.} {\bf 68#1}}
\def\jmp#1{{J.Math.Phys.}}
\def\jaat{J.Astrophys.Aerosp.Technol.\ {}} {}
\def \epj#1{{Eur.Phys.J.} {\bf C#1}} 
\def \jcap{JCAP\ {}}{}

\end{document}